# Long-Term Load Forecasting Considering Volatility Using Multiplicative Error Model

Swasti R. Khuntia, Jose L. Rueda, Mart A.M.M. van der Meijden

**Abstract**

Long-term load forecasting plays a vital role for utilities and planners in terms of grid development and expansion planning. An overestimate of long-term electricity load will result in substantial wasted investment in the construction of excess power facilities, while an underestimate of future load will result in insufficient generation and unmet demand. This paper presents first-of-its-kind approach to use multiplicative error model (MEM) in forecasting load for long-term horizon. MEM originates from the structure of autoregressive conditional heteroscedasticity (ARCH) model where conditional variance is dynamically parameterized and it multiplicatively interacts with an innovation term of time-series. Historical load data, accessed from a U.S. regional transmission operator, and recession data for years 1993-2016 is used in this study. The superiority of considering volatility is proven by out-of-sample forecast results as well as directional accuracy during the great economic recession of 2008. To incorporate future volatility, backtesting of MEM model is performed. Two performance indicators used to assess the proposed model are mean absolute percentage error (for both in-sample model fit and out-of-sample forecasts) and directional accuracy.

## 1. Introduction

Load forecasting in long-term horizon is important for electric utilities and planners in terms of grid expansion planning, future investments and revenue analysis for long-term decision making process. Moreover, it plays a vital role in economic and social development of a country (or specific region in case of some utilities). A more realistic range of future generation scenarios can be modeled when the electricity consumption is increasing at a faster rate in this globalizing world. For instance, annual load forecasting is favored among utilities and is one of the common long-term load forecasting problems. It can alleviate the disparity between demand and generation, thereby maintaining the required level of security of supply. Choosing a right horizon for long-term varies from one utility to another based on their policies. Usually a monthly or yearly time-step for one to ten years ahead in long-term load forecasting is helpful in inter-tie tariff setting and long-term grid investment return problems.

It is often difficult to forecast load over a such a long planning horizon and it is due to the stochastic nature of the forecasting process and influential parameters. Most of these parameters are, by nature, unpredictable and uncontrollable. Examples are socio-economic developments, occurrence of special events and/or climatic conditions, and regulatory requirements. Any considerable deviation in forecast results in over expenditure on generation/transmission infrastructure or energy resource waste. Hence, in order to improve the forecast accuracy in long-term horizon, attention is needed either in terms of *improvement of existing employed techniques* or *development of a new technique to consider all the aforementioned factors*. Forecast accuracy influences the favor of generation and transmission companies on their plans to combat future load growth and market volatilities. Based on the forecast, electric utilities coordinate their resources to meet the actual demand using a cost-effective plan. Fig. 1 depicts the future complexity in context of load forecast, various players in action and the inter-dependency that needs attention too. Stochasticity in future scenarios, energy users and the uncertainty associated with evolution of prosumers complicates the forecasting methodology. Not to be forgotten is the spatial complexity as the area expands from distribution system operator (DSO) level to multi-transmission system operator (TSO) level.

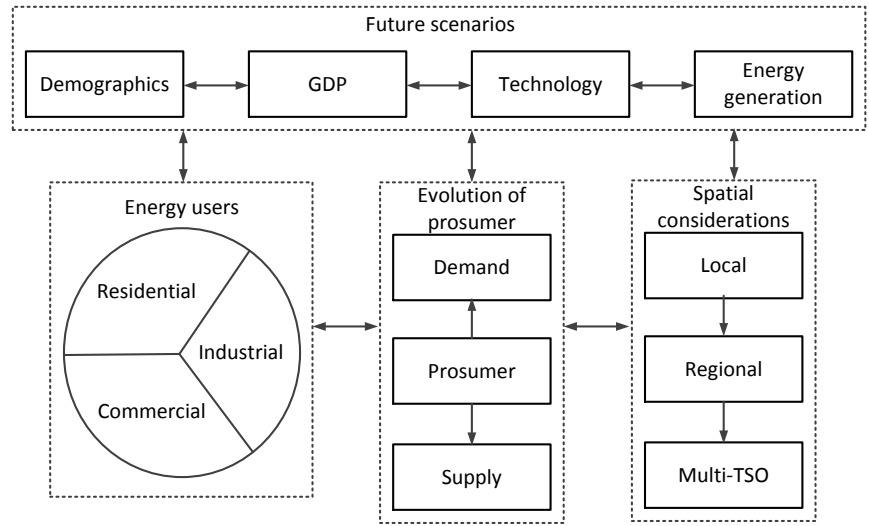

**Fig. 1: Complexity in long-term load forecast [GDP: Gross Domestic Product, DSO: Distribution System Operator, TSO: Transmission System Operator]**

This research sums up the need for accurate load forecasting in long-term horizon as,

- Firstly, moving towards greener future is accredited with development in new technology and integration of renewable energy into primary grid while discarding fossil fuels is becoming important. In the Paris Agreement 2016 [1], it was agreed upon to move towards renewable energy from the more conventional energy. Such a move is realized with accurate and reliable forecast of the electrical energy demand. Despite advancements in battery technology, storing energy for long-term purpose is not the viable option. Thus, accurate and reliable forecasting is required for planning the right tools.

- Secondly, considerable changes in weather factors like temperature, rainfall and hot/cold days. Any change in climatic variables will have a direct impact on the demand pattern. Erratic weather events posed due to climate change pose some serious burden on forecasters to accurately model load growth considering long-term horizon.

- Thirdly, maintaining security of supply during the energy transition. In today's date, existing grids are performing under stress to deliver the growing demand in the presence of variable stochastic renewable energy sources.

- Lastly, black swan events like the great economic recession of 2008 jolted the economic backbone of many countries. Its effect was widespread and energy investment worldwide plunged into tougher financing environment and weakening final demand for energy [2]. This reminds the importance to study the financial aspects of long-term forecasting by energy forecasters in electric utilities.

Based on the needs, the key contributions of this research can be listed as:

- Presents recent advancement in long-term load forecasting in terms of techniques and models developed.
- Provides a comprehensive and critical evaluation of long-term load forecasting considering volatility.
- Use of multiplicative error model (MEM) to model conditional mean and forecasting aggregated zonal load. In this research, we consider a forecast horizon of 4 years as a solution for electric utilities and planners based on the fact that construction of offshore wind farms takes approximately 3-4 years depending on the capacity [3].

The rest of the section is organized as follows: section 2 gives a background on long-term load forecast and the need for accounting volatility, section 3 introduces multiplicative error model and followed by forecast methodology in section 4. Section 5 analyses the forecast results based on real data. Finally, section 6 concludes the work.

## 2. Background on Long-term Load Forecast

Electric load forecasting in long-term horizon is an important part of transformation of electric power systems and it has appealed more and more attention from both academic and industry. By principle, a load forecasting model aims at mathematical representation of the relationship between load and influential parameters. Such a model is identified with coefficients that are used to forecast the future values by extrapolating the relationship to a desired lead time. Eventually, the accuracy of model depends on both the selected model as well as accuracy of the estimated parameters. Literature study reveals that long-term load forecasting received less attention compared to short-term load forecasting. It is because of the complexity involved in achieving accurate forecasts. Long-term load forecasting is based on the integration of concepts from theoretical foundations of economic theory with knowledge on financial, statistical, probability, and applied mathematics to make inferences about the load growth/fall and technology evolution. Ref. [4] illustrates rationally the concept of long-term load forecasting and also presents recent development within electric power industry. Ref. [5] performed a study on past, current and future trends in energy forecasting while stating the trend in spatial, short-term and long-term load forecasting, and energy price forecasting in a lucid manner. Ref. [6] proposed three methods suitable for long-term forecasting as: time-series approach, econometric approach, and end-use approach. For long-term forecasting, all approaches require historical data and they are broadly categorized into traditional (or statistical) and non-traditional (or artificial intelligence AI) based methodologies. Still it is widely accepted and it is proven by literature study.

Traditional methods include regression-based model and time series methods. Ref. [7] proposed univariate autoregressive models to forecast load with monthly time-step in Lebanon. Multiple linear regression models was proposed in [8]. Ref. [9] implemented a knowledge-based expert system to support the choice of the most suitable load forecasting model with practical application. However, traditional methods are criticized for their weakness of non-linear fitting capability. In AI-based techniques, artificial neural network (ANN) is one of the most popular model. Its application in forecasting Greek long-term energy consumption for the years ahead is reported in [10]. Ref. [11] used ANN on the Egyptian electrical network for long-term peak load forecasting. Ref. [12] reported the superiority of ANN for medium and long-term load forecasting in terms of accuracy and robustness. Hybrid of fuzzy and ANN are reported in [13] for forecasting Taiwan's annual electricity load and in [14] for long-term electrical energy consumption in India. Other AI techniques include support vector regression models (SVR) [15-16] and SVR with simulated annealing algorithms [17].

Use of metaheuristic methods such as genetic programming [18], fruit-fly algorithm [19], gravitational search algorithm [20] and particle swarm optimization (PSO) [21-22] are also reported. Other methods include long-term forecasting based on partial least squares method [23] and complete decomposition method [24]. Recent study includes forecasting for country specific such as Spain [25], Greece [10,26], Lebanon [7], Turkey [27-29]. More recently, ref. [30] used gene expression programming for long-term prediction of electrical energy consumption in ASEAN-5 countries and projected up to 2030 according to rolling-based forecasting procedure. The results are compared with those obtained from ANN, SVR, adaptive neuro-fuzzy inference system (ANFIS), rule-based data mining algorithm, gene expression programming (GEP), linear and quadratic models optimized by PSO, cuckoo search algorithm (CSA) and backtracking search algorithm (BSA).

It is evident that the developed models aim at predicting accurate peak load or electrical energy consumption while comparing with any classical model. However, one aspect that has received less attention in long-term load forecasting when the whole energy scenario is growing in terms of complexity and dynamics is volatility. The concept of volatility is prevalent in financial markets and it refers to the degree of erratic variations of a process over time. It is used as a criterion to study the risk associated with a financial asset. Ref. [31] showed that power markets have greater volatility levels than other financial markets like crude oil, natural gas or stock prices. Literature study reveals volatility studies on various electricity markets: Spanish, Californian, UK and PJM electricity markets [32], Ontario and some of its neighboring markets [31], German market [33], Australian electricity market [34] to name a few. Ref. [35] examined and compared the volatility of 14 deregulated markets through the "price velocity" metric. Nordic pool was studied in [36] considering volatility clustering, log-normal distribution and long-range correlations. In time-series forecasting of electrical load, volatility is defined as a deviation from the mean which corresponds to risk. An advantage of such an approach is that once the time-series model is understood, it is possible to simulate the data generation for any lead time in future. Ref. [37] explained the importance of volatility in long-term load forecast, which no work reported earlier. Extending the concept of volatility forecast to load forecast in long-term horizon is adopted in this research.

Volatility is a fundamental issue in financial and econometrics domain, and virtually present in all financial decision making. In fact, volatility is forecastable because of a number of persistent properties: (i) it appears in clusters, (ii) it changes over time and has unusual jumps, (iii) it does not grow to infinity and is persistent in specific time-span, and (iv) it reacts different for an increase or decrease of the considered entity. For instance, load forecast in long-term horizon takes into account socio-economic factors like population growth and gross domestic product (GDP) along with explicit factors like historical load and weather data. Presence of economic factors induces volatility, or what is called as implied volatility. In fact, implied volatility is generally treated to be the best available forecast as it has certain characteristics that can increase the accuracy of forecast values. Likewise load, future volatility prediction is an extremely difficult task because the actual realization of the future process volatility will be influenced by events that happen in the future. Thus, it is important to develop a model that can fit the sequence of calm and turbulent periods. Studies reveal that ARIMA technique, one of the most well-known forecasting techniques, is inadequate in long-term forecasting task because it suffers from mean convergence problem. To address the short-coming and treating volatility as influential parameter, the next section introduces the concept of MEM and its application to load forecast in long-term horizon.

## 3. Multiplicative error model for long-term load forecast

Multiplicative error model (MEM) was introduced by Engle in 2002 [38] as an adaptation of autoregressive conditional duration model [39] to be used for time-series that always receive positive values. Literature study on MEM reveals its application in financial risk and volatility forecasting [40-42]. A search about application of MEM in load forecasting reveals no information, not even for short-term forecasting which is common among forecasters. Hence, the proposed model is first-of-its-kind to introduce MEM for load forecasting. As electric load is always represented as a non-negative time-series, MEM is presumed to be a good fit to forecast. The MEM for a non-negative time-series $(y_t)$ on $[0, +\infty)$ and considering $\mathcal{F}_{t-1}$ as information available for forecasting $y_t$ is written as [38]:

$$y_t = \mu_t \varepsilon_t \tag{1}$$

where, the range of the disturbance $\varepsilon_t$ is non-negative on $[0, +\infty)$, unit mean and unknown constant variance given as $\varepsilon_t | \mathcal{F}_{t-1} \sim D(1, \psi)$ for positive distribution $D$. $\mu_t$ is conditional on $\mathcal{F}_{t-1}$ and positive, described on a parameter vector $\theta$ as:

$$\mu_t = \mu(\theta, \mathcal{F}_{t-1}) \tag{2}$$

When $\mathcal{F}_{t-1}$ includes only historical values of the series, $\mu_t$ can be generalized as:

$$\mu_t = \delta + \sum_{i=1}^{p} \alpha_i \mu_{t-i} + \sum_{j=1}^{q} \beta_j y_{t-j} \tag{3}$$

where, $\delta$ is constant, term $\sum_{i=1}^{p} \alpha_i \mu_{t-i}$ represent an inertial component, and term $\sum_{j=1}^{q} \beta_j y_{t-j}$ represent more recent observation. Equation (3) is referred to as referenced MEM of order $(p, q)$. Model specifications can be modified to adapt to the needs of load forecast. For instance, residuals at $t$-th observation denoted as $\vartheta_t = y_t - \mu_t$ and $\alpha_1^* = \alpha_1 + \beta_1$, equation (3) can be written as:

$$y_t = \delta + \alpha_1^* y_{t-1} + \vartheta_t - \alpha_1 \vartheta_{t-1} \tag{4}$$

Equation (4) represents an ARMA model with heteroskedastic errors, and is the cornerstone of this modeling approach. The procedure of finding and validating a suitable MEM for a given dataset is discussed in next section.

## 4. Forecast Methodology Considering Real Data

Since In order to realize a suitable long-term forecasting model, one must start with a rich historical database, construct the model, identify the appropriate model and finally evaluate the forecast results. Fig. 2 shows the steps to forecast load using MEM. Since MEM falls under time-series models, we follow the Box-Jenkins methodology of building model with certain adaptations [43]. Starting from data preparation, the first part involves stationarity checking, data fitting and model identification while checking various statistical properties of the time-series. Identifying the right model, estimating parameters and checking the model adequacy falls under this part. In the second part, MEM is validated for forecasting both as in-sample fit and out-of-sample prediction. Modeling of MEM starts with identification of autoregressive and moving average parameters of non-negative time-series that has predictive power regarding the directional change, and later added by persistent error specifications that eventually improves forecasts. MEM differs from linear regression models in the sense that the mean equation, which is a scalar factor, is multiplied with the independent and identically distributed ($i.i.d.$) error term. The scalar factor evolves in a conditionally autoregressive manner, hence, favorable for forecasting. The assumption of $i.i.d.$ means that the error terms behave randomly with constant mean and variance over a considered time-horizon. However, in reality, both the load and error time-series are highly correlated and do not behave as an $i.i.d.$ process which is realized in this methodology.

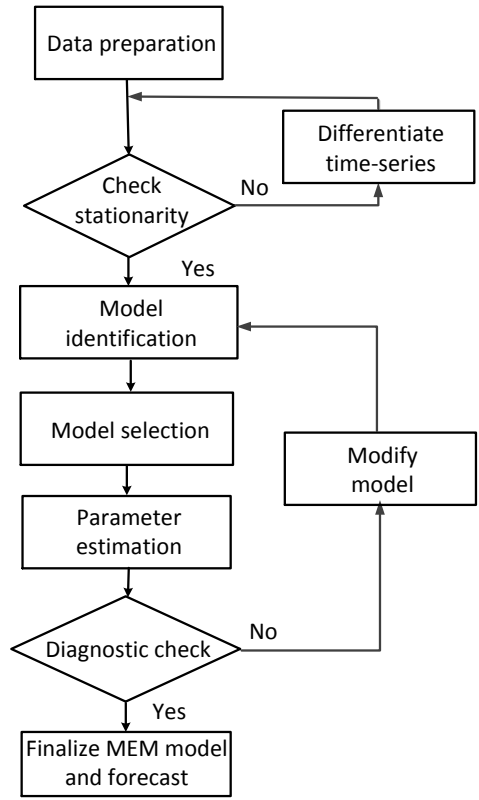

**Fig. 2: Flowchart for model identification, selection and forecast**

### A. Database generation

Forecast accuracy strongly depends on the quality of available historical data. A poor history, composed only by anomalous or average events, may polarize the analysis and affect the quality of the forecast values. For this study, historical data of a specific load zone region under a U.S. regional transmission operator [44] and data describing economic recession as extracted from National Bureau of Economic Research (NBER) [45] is considered. Hourly load data for years 1993-2016 is extracted and sampled to monthly aggregated load as shown in Fig. 3. The use of monthly time-stamp enables in understanding the monthly energy consumption. Recession data for years 1993-2011 is used to build the predictor. From Fig. 3, it seems that the load growth is on fairly positive side apart from few incidents where a downturn in demand is observed. Such an incident is the year span of 2006-2009, where year 2007-2008 is identified by large variability in demand value because of spikes and negative demand growth coincide with the great recession of 2008.

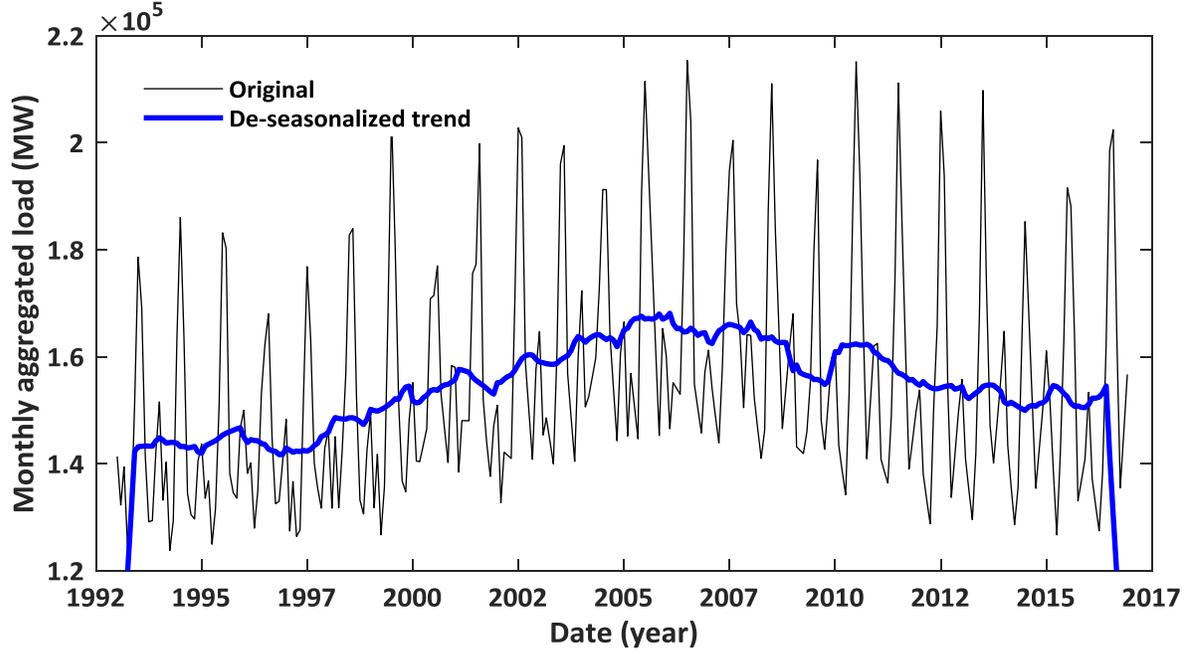

**Fig. 3: Monthly aggregated load data and de-seasonalized trend for years 1993-2016**

### B. Stationarity and autocorrelation test

A visual inspection of Fig. 3 suggests non-stationary time-series with a linear trend and seasonal periodicity. Test reveals that non-stationarity is apparent as both mean and variance increase with time. The class of MEM requires time-series to be stationary so that its statistical (up to the second order moment) properties do not depend on time. This is coherent with any time-series forecasting because non-stationary time-series are erratic and unpredictable. Phillips-Perron (PP) test is used for stationarity check [46]. For any time-series $x_t = ax_{t-1} + e_t$ where $e_t$ is the residual, PP test checks for null hypothesis ($H_0: a = 0$ vs. $H_1: a \neq 0$). Use of PP test is preferred over the widely used augmented Dickey-Fuller (ADF) because of its non-parametric nature. In addition to the steps from DF test, PP test corrects the statistics to account for autocorrelations and heteroscedasticity. The time-series is checked for 0 lags and both the tests reject the null hypothesis with $p$-value of 0.001. Thus, the time-series is differenced to obtain a stationary time-series and next step is to determine the presence of autoregressive or moving average terms to correct any autocorrelation that exists in the differenced time-series.

Two tests used to check null hypothesis ($H_0: no \ autocorrelation$ vs. $H_1: correlation$). are Ljung-Box Q-test ($Q$) and Durbin-Watson ($D$) test [47]. The $Q$-test statistic for $R$ residuals, $L$ lags is written as,

$$Q = R(R + 2) \sum_{l=1}^{L} \left( \frac{\rho(l)^2}{(R - l)} \right) \tag{5}$$

where, $\rho(l)$ is the autocorrelation coefficient at lag $l$. The Durbin-Watson statistic ($D$) is conditioned on the order of the observations (rows) or the number of months in our study. The $D$-test statistic for $n$-observations is written as:

$$D = \frac{\sum_{i=2}^{n}(R_i - R_{i-1})^2}{\sum_{i=1}^{n} R_i^2} \qquad (6)$$

Presence of serial correlation in a time-series indicates that the values of adjacent observations are correlated. Fig. 4 shows the autocorrelation and partial autocorrelation plot giving evidence of presence of autoregressive and moving average parameters. The ACF plot reveals the presence of significantly large autocorrelations, particularly at every 12th lag. Presence of autocorrelation suggests the data is dependent and correlated and needs modification. Table 1 displays the detailed statistics of original and differenced time-series.

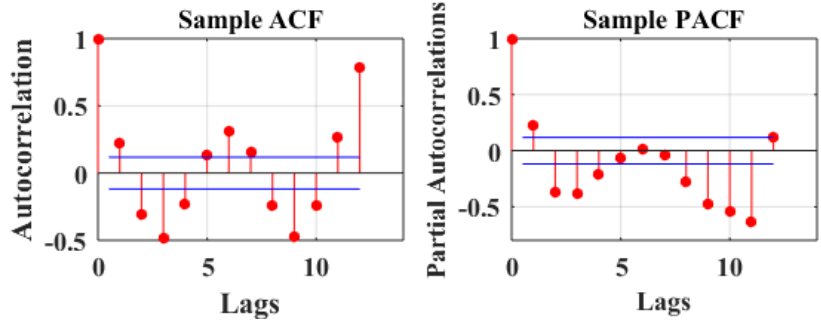

**Fig. 4: Sample autocorrelation function (ACF) and partial autocorrelation function (PACF) plot of differenced time-series**

**Table 1: Detailed Statistics of Load Time-series**

|  | Original | Residuals |
| --- | --- | --- |
| Mean | 154581.1 | 53.40888 |
| Max | 215379.9 | 45842.8 |
| Min | 123786.5 | -49286.9 |
| Median | 149928.9 | 433.1667 |
| Standard deviation | 20514.4 | 17879.6 |
| Skewness | 1.015012 | -0.10097 |
| Excess kurtosis | 3.449577 | 3.005507 |

### C. Volatility check and Multiplicative error modeling

Next step in modeling is to check if the differenced time-series shows any cluster of volatility and satisfy the homoscedastic assumption of constant variance or heteroskedastic behavior. It may happen that squared values of the differenced time-series exhibit significant serial correlation. It means that values are again dependent but serially uncorrelated. So, the sample autocorrelation and partial autocorrelation test is repeated for squared residual followed by Q-test and DW-test. The tests re-confirm our model selection [48], and the corresponding plot of autocorrelation and partial autocorrelation function is shown in Fig. 5.

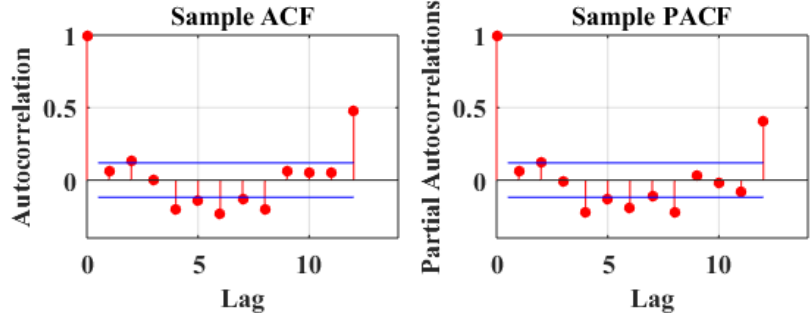

**Fig. 5: Sample autocorrelation function (ACF) and partial autocorrelation function (PACF) plot of squared residuals**

The ACF and PACF plots in Fig. 5 verifies the presence of conditional heteroscedasticity and also facilitates in identifying the appropriate order for de-seasonalized differenced time-series. As stated in ref. [38], generalized autoregressive conditional heteroscedasticity (GARCH) models are a form of MEM and is used in this study. With reference to equation (1), if $\mu_t$ is the conditional expectation of $y_t$, its parameters can be estimated by specifying a GARCH for the conditional second moment of $\sqrt{y_t}$ while imposing its conditional mean to be zero. Ref. [49] augmented the regression model with GARCH error modeling, and the same concept is adapted for this study. The standard model common to both the processes and its square while rewriting equation (1) is:

$$Z_t = \sqrt{h_t} e_t$$

$$Z_t^2 = h_t e_t^2$$

(7)

In the squared equation, the dependent variable ($Z_t$) is non-negative with mean $h$ and a non-negative multiplicative $e_t \sim i.i.d. (0,1)$ with unit mean. This can be estimated by taking the load residual as the dependent variable of a GARCH model. The GARCH model is an extension of the ARCH model, in the way that it allows current volatility to be dependent on its lagged values directly. For more information on ARCH and GARCH, ref. [50] is recommended. The model can be estimated by taking $Z_t$ as the dependent variable, with specifications of zero mean and an error process. In such case, the *conditional variance* is then the *conditional mean* of $Z_t^2$ [51]. Rewriting equation (1), the GARCH model with order $p \geq 0$ and $q \geq 0$ is defined as [50]:

$$Z_t = \sqrt{h_t} e_t$$

(8)

$$h_t = \alpha_0 + \sum_{i=1}^{p} \alpha_i Z_{t-i}^2 + \sum_{j=1}^{q} \beta_j h_{t-j}^2$$

(9)

for the square root of duration, and where $\alpha_0 > 1$, $\alpha_i \geq 0$ and $\beta_j \geq 0$ are constants with

$$\sum_{i=1}^{p} \alpha_i + \sum_{j=1}^{q} \beta_j < 1$$

(10)

and $e(t)$ is independent of $Z_{t-k}, k \geq 1$.

Selecting the right order $(p, q)$ is achieved by following one of the many order selection tests. Akaike Information Criteria (AIC) and Bayesian Information Criteria (BIC) tests are chosen in this study. The reasons for choosing the two criterions are that both the tests assess the fit between model predicted and original values and penalize models with larger number of parameters. Tests confirmed the use of order $(1,1)$ multiplicative error model. Fig. 6 shows the innovation plot for a sample size of 101 (0-80 range shown in Fig. 6), and it can be concluded that clusters of volatility appear in a periodic manner. Thus, the movement of non-linearity is not only dependent on the previous values but for the whole time-series it is uncorrelated. Volatility tends to cluster into periods with higher and lower volatility. This effect proves that volatility at some time must be dependent on its historical values, say with some degree of dependence.

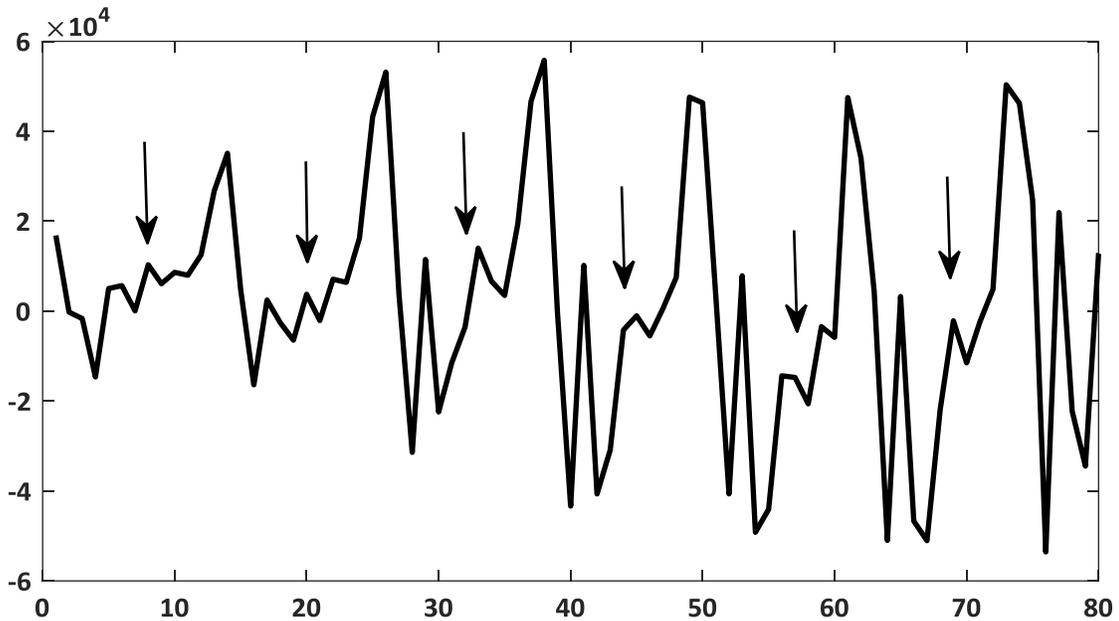

**Fig. 6: Conditional variance showing clusters of volatility (pointers show the clusters)**

After the model order is identified, maximum likelihood estimator is used to estimate the parameters. Regardless of the low standard errors, parameter estimation is still feasible. As the sample size runs from $N \to \infty$, the probability that the value of the estimators shows a large divergence from the true (which is unknown) parameter values goes to 0, making it a consistent estimators. Estimation is achieved with conditional variance $h_t \sim i.i.d.(0,1)$, and with an assumption that error distribution follows student t-distribution, a version of the generalized error distribution, whose density is given as,

$$f(x) = \frac{v e^{\left(-\frac{1}{2}\left|\frac{x}{\lambda \mu}\right|^v\right)\frac{1}{\mu}}}{\lambda 2^{\left(1+\frac{1}{v}\right)}\Gamma\left(\frac{1}{v}\right)} \tag{11}$$

where $v$ is positive measure of fat tail, $\lambda = \sqrt{2^{-(2/v)} \frac{\Gamma(1/v)}{\Gamma(3/v)}}$, and $\Gamma(.)$ is the gamma function defined as $\Gamma(x) = \int_0^\infty y^{x-1} e^{-y} dy$. This assumption helps in better modeling of excess kurtosis (in Table 1). It also approximates the normal distribution as the degrees of freedom grow to infinity. Presence of fat tail is evident from Q-Q plot in Fig. 7.

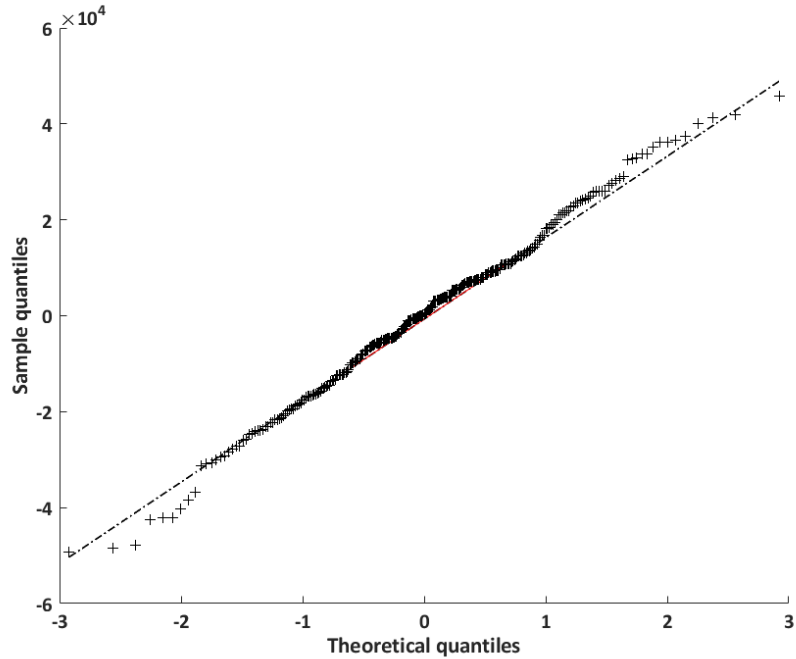

**Fig. 7: Q-Q plot of residuals**

## 5. Result analysis

Result analysis consists of two parts: first part consist of in-sample model fit using load and economic data followed by out-of-sample forecast, and the second part is checking directional accuracy by forecasting for the year 2008 during the great economic recession.

### A. In-sample model fit and out-of-sample forecast

A forecast horizon of 4 years is chosen in this study. It is based on the assumption that off-shore wind-farm plant needs at least 3-4 years for completion, which is itself a long-term grid development action [3]. For the in-sample model fit, the study embodies fitting the MEM using load data and recession data of years 1993-2011, and then evaluating its performance on the test data set for years 2013- 2016. When assessing point forecasts with mean square errors, it appears to be useful to use a longer in-sample period for model estimation as followed in this study. Fig. 8 shows the in-sample forecast for the years 2013-2016. A more closer look at the forecast results reveals the superior fit of MEM. The model performance is calculated with Mean Absolute Percentage Error ($MAPE$) metric. For two sets of $n$-observations ($x_{i,...,n}, y_{i,...,n}$), $MAPE$ is defined as

$$MAPE = \frac{100}{n} \sum_{i=1}^{n} \left| \frac{y_i - x_i}{y_i} \right| \qquad (12)$$

where $y_i$ is the original monthly aggregated load and $x_i$ is the predicted monthly aggregated load. In-sample model fit accuracy is achieved with MAPE of 5.68%.

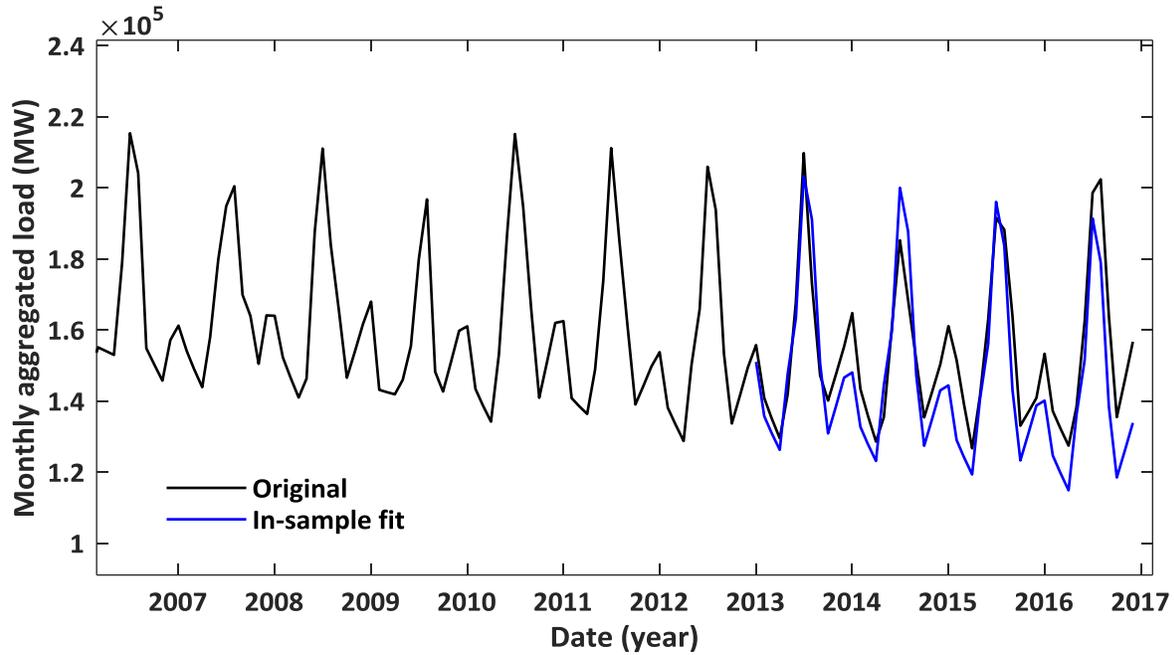

**Fig. 8: In-sample fit for years 2013-2016 with a MAPE of 5.68%**

No forecasting analysis is complete without performing out-of-sample forecasts. For better out-of-sample forecasts, the most crucial choice is splitting the series between training and test periods. Unfortunately, no study exists so far that discusses on how to choose the decision point [52]. In this study, training dataset of years 1993-2012 is chosen to forecast the next four years (2013-2016). The accuracy of the MEM model is improved with backtesting technique where the aim is to achieve a dynamic model that can address the future volatilities. With 48 months as forecast horizon and monthly timestamp, the MEM is build every month and forecasts ahead 48 months. The forecast result compares with original values and averages the error. In such a manner, the out-of-sample result improves as the model learns and adapts from past results. Fig. 9 shows the out-of-sample forecast results with MAPE of 7.09%. A high error percentage as compared in-sample model fit is understood from the long forecast horizon.

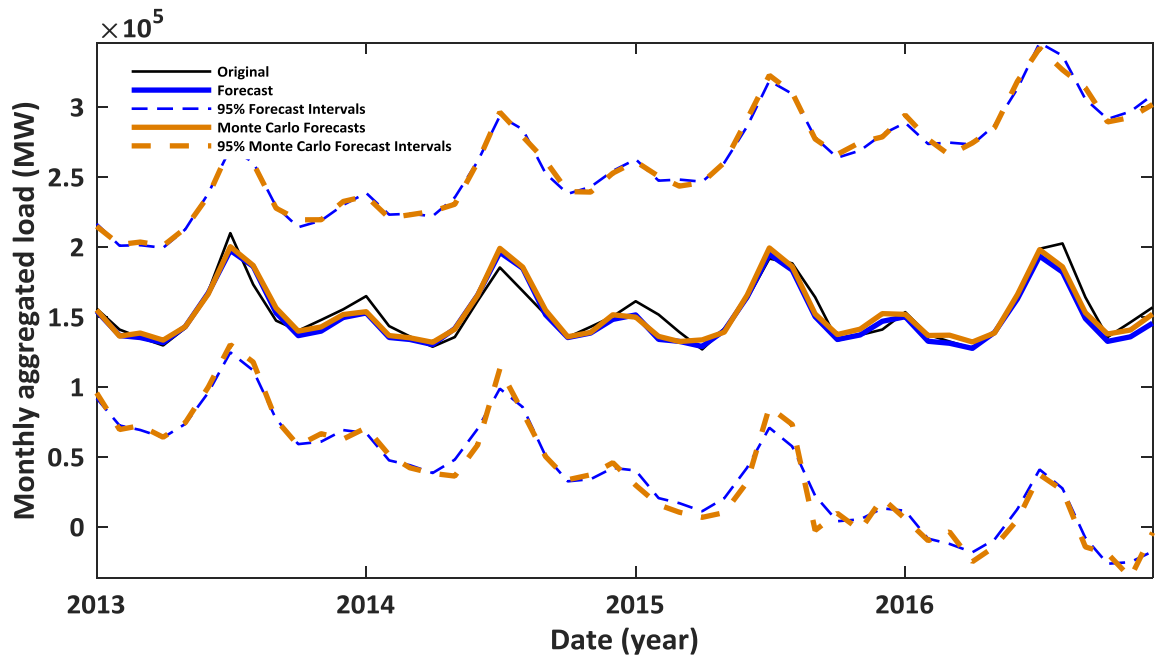

**Fig. 9:** Out-of-sample forecast for years 2013-2016 with 95% confidence interval both upper and lower and a MAPE of 7.09%

To better evaluate the model accuracy, Monte Carlo simulation is run for 500 sample paths by choosing a confidence interval of 95%. The motivation behind calculating range forecasts this way is based on the conclusion that, while forecasters tend to be overconfident in forecasting confidence intervals, they are much more accurate in evaluating other forecasters' confidence intervals and estimating the likelihood that a particular forecast will be accurate. In this way, the values within the confidence interval of the conditional mean describe the considerable range of values of the point on the line. Thus, the conditional mean for all values of time-series indicate how much the entire MEM prediction can considerably move from sample to sample. It eases in predicting the range of likelihood values that an observation in the next time step may take. The confidence interval of the out-of-sample forecast presents a range for the mean rather than the distribution of individual data points. Fig. 9 shows a comparative analysis of out-of-sample forecast and the Monte Carlo simulation results. Both the forecast as well as confidence intervals from the two outputs are virtually indistinguishable. To understand the intervals, a value of 0.05 corresponds to predicted upper and lower intervals where there is a 5% chance that original values will not be in that range.

While evaluating the forecast results, we take a glimpse back at Fig. 7 and observe that the data is skewed to the right. The Q-Q plot also display sizeable excess kurtosis or fat tails. Also referring to Table 1, the high skewness and kurtosis value is an indicator of non-normal time-series. To verify the claim, Jarque–Bera (JB) test is considered in our study. It is usually used for large data sets, because other normality tests are not reliable for large data sets. The JB-test verifies the null hypothesis ($H_0$: $normal$ vs. $H_1$: $non-normal\ distribution$). The JB-test statistic is written as [53]:

$$JB = N\left(\frac{s^2}{6} + \frac{(k-3)^2}{24}\right)$$

(13)

where, $N$ is the sample size, $s$ is the skewness coefficient and $k$ is the kurtosis coefficient. A value of 1 from JB-test indicates the data is non-normally distributed. The residual distribution is fitted with Student's t-distribution, which has a thicker tail and is thus more tolerant to extremes. The study is repeated by including both fat-tails and volatility to verify if the forecast improves and the result is significant. Fig. 10 shows the forecast for years 2013-2016 with and without accounting for fat-tails. Inclusion of fat-tail is significant because it represents a greater likelihood of extreme events occurring similar to the financial crisis, also called the black swan event [54]. Some notable features of volatility that should be clearly mentioned are: volatility appears in clusters apparent from Fig. 7, volatility changes over time and that jumps in the volatility are unusual, volatility does not grow to infinity; it rather stays within some spans, and the fourth characteristic is that the volatility reacts different on a drop in the demand than it does for an increase in the demand. The estimated MEM parameters are shown in Table 2. To support the range for in-sample model fit, one of the assumptions in the study is that a *t-statistic > 2* in magnitude correspond to approximately a 95% confidence level. The t-statistic column is the parameter value divided by the standard error, and is normally distributed for large samples. It measures the number of standard deviations the parameter estimate is away from zero.

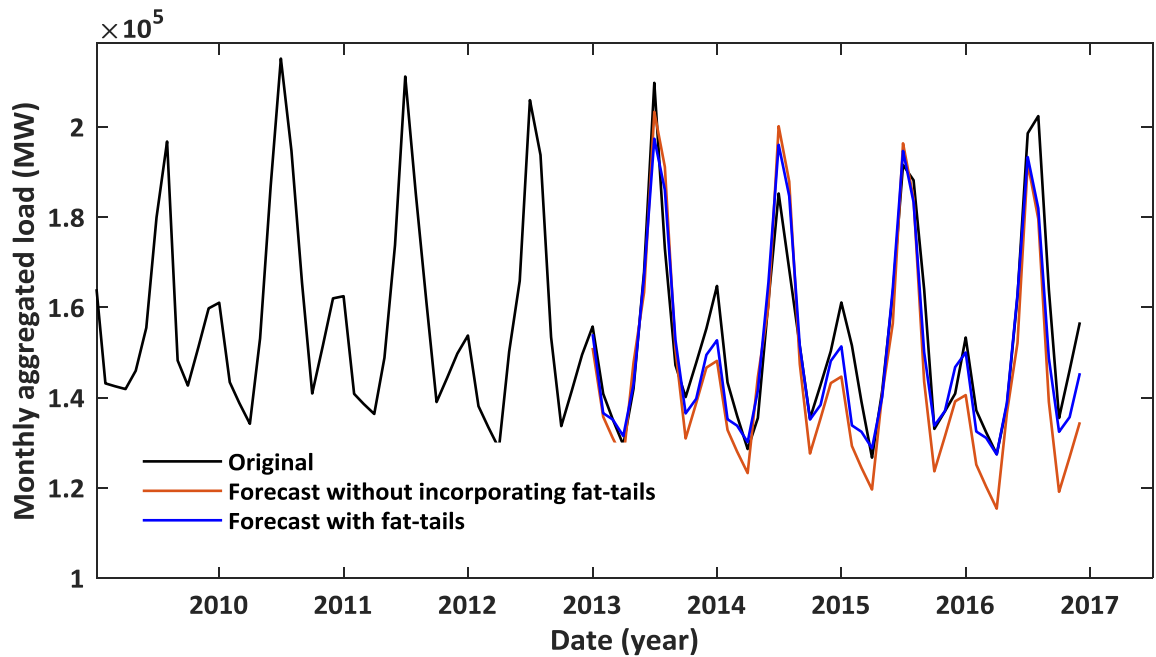

**Fig. 10: Forecast for years 2013-2016 with and without incorporating fat-tails**

**Table 2: Multiplicative Error Model Parameters**

| Parameter | Multiplicative error model values | Standard errors | t-statistic |
|-----------|-----------------------------------|-----------------|-------------|
| $\alpha$ | 1.368271e+07 | 0.000153268 | 7.22314e+10 |
| $\alpha_1$ | 0.703561 | 0.012491 | 40.1238 |
| $\beta_1$ | 0.0241376 | 0.0201932 | 1.5713 |

### B. Directional accuracy of forecast methodology

The second part of result analysis is checking directional accuracy during the year 2008 when the great economic recession hit the whole world and U.S. was largely affected. Since the data is from U.S. utility, it was decided to check the robustness of model during that period. An out-of-sample forecast is performed for the year 2008 with a training dataset of years 1993-2007. When economic factors play a pivotal role, the need to study market movements is important. Not many forecast studies include the significance of directional forecasting and how its accuracy supports the statistical parameters. Fig. 11 shows two overlapped time-series. A long period of uniform load growth was interrupted in early 2000s till mid-2000s. In fact, the 2000s show two distinct jumps in historical load data (seen in Fig. 3): one was triggered by energy crisis because of fluctuating oil prices, and one was prompted by the great recession of 2008. Since then, load growth has regularly displayed volatility relative to the pre-2000s. As the real load growth have not changed much over time, still large fluctuations tend to be concentrated over somewhat short periods, thus embodying directional accuracy along with improved and accurate forecast result is preferred.

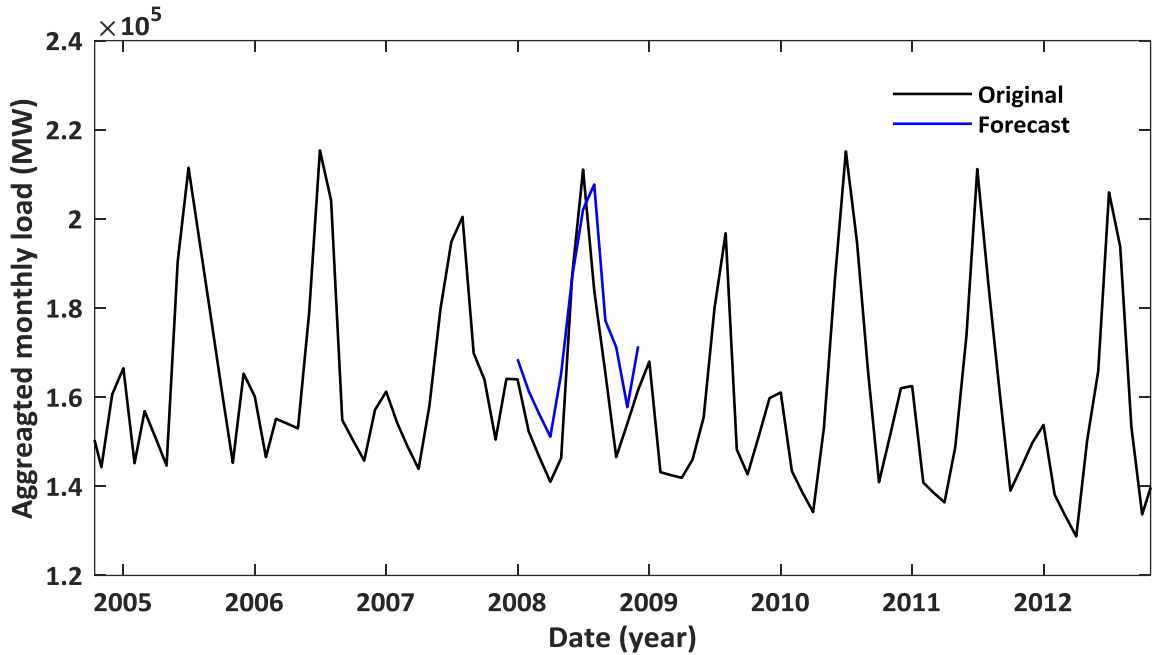

Fig. 11: Forecast during great recession of 2008

### 6. Conclusion

In this study, a first-of-its-kind implementation of MEM to forecast load in long-term horizon is presented. It aimed at presenting not only an accurate long-term forecast methodology but also a robust methodology that can withstand the volatility. Advantage of MEM over other time-series models is the ease of modeling volatility. The term conditional variance in MEM denotes the dependency on past sequence of events and is quite contrasting to unconditional which implies long-term behavior assuming null knowledge on past events. In this study, consideration of conditional variance has resulted in improved forecast for long-term horizon. Though both conditions are vital in volatility forecasting, usage of conditional variance in MEM outperforms.

The two performance indicators for this study are: point forecast with a low error percentage as proved by mean error metrics for both in-sample and out-of-sample forecasts, and the directional accuracy during recession. The inclusion of heteroscedastic errors that improves forecast performance and also shows that it is possible to predict the direction of change of residuals in the presence of conditional heteroscedasticity, even if the residuals themselves cannot be predicted.

MEMs inherit many properties from the theory of ARMA models. Some of them being: volatility clustering, fat-tails and mean reversion, which helps in learning the statistics of volatility in time-series. It is also learnt that the time-series exhibit skewness and excess kurtosis, well supported by Jarque-Bera test. The two properties are considered by adopting Student's t-distribution in addition to the normal distribution in an effort to improve forecast accuracy.

## 7. Acknowledgement

The research leading to this result has received funding from the European Union Seventh Framework Programme (FP7/2007-2013) under grant agreement No. 608540 GARPUR project http://www.garpur-project.eu. The scientific responsibility rests with the authors.

## 8. References

[1]     Weblink: http://unfccc.int/paris_agreement/items/9485.php [Accessed on 20th July, 2018]

[2]     Weblink: https://www.iea.org/publications/freepublications/publication/impact.pdf [Accessed on 20th July, 2018]

[3]     S.R. Khuntia, B.W. Tuinema, J.L. Rueda, and M.A.M.M. van der Meijden, "Time-horizons in the planning and operation of transmission networks: an overview," *IET Gen. Trans. Distr.*, vol. 10, no. 4, pp. 841-848, 2016.

[4]     S.R. Khuntia, J.L. Rueda, and M.A.M.M. van der Meijden, "Forecasting the load of electrical power systems in mid-and long-term horizons: a review," *IET Gen. Trans. Distr.*, vol. 10, no. 16, pp. 3971-3977, 2016.

[5]     T. Hong, "Energy forecasting: Past, present, and future," *Foresight: The International Journal of Appl. Forecast.*, vol. 32, pp. 43-48, 2014.

[6]     E.A. Feinberg and D. Genethliou, *Load Forecasting*. New York, NY, USA: Springer Science, ch. 12, 2005.

[7]     S. Saab, E. Badr and G. Nasr, "Univariate modeling and forecasting of energy consumption: the case of electricity in Lebanon," *Energy*, vol. 26, pp. 1-14, 2001.

[8]     Z. Mohamed and P. Bodger, "Forecasting electricity consumption in New Zealand using economic and demographic variables," *Energy*, vol. 30, pp. 1833-1843, 2005.

[9]     M.S. Kandil, S.M. El-Debeiky, N.E. Hasanien, "The implementation of long-term forecasting strategies using a knowledge-based expert system: Part-II," *Elect. Power Syst. Res.*, vol. 58, no. 1, pp. 19-25, 2001.

[10]    L. Ekonomou, "Greek long-term energy consumption prediction using artificial neural networks," *Energy*, vol. 35, no. 2, pp. 512-517, 2010.

[11] A.A. Abou El-Ela, A.A. El-Zeftawy, S.M. Allam, G.M. Atta, "Long-term load forecasting and economical operation of wind farms for Egyptian electrical network," *Elect. Power Syst. Res.*, vol. 79, no. 7, pp. 1032-1037, 2009.

[12] C. Xia, J. Wang and K. McMenemy, "Short, medium and long term load forecasting model and virtual load forecaster based on radial basis function neural networks," *Electr. Power Energy Syst.*, vol. 32, no. 7, pp. 743-750, 2010.

[13] T. Chen, "A collaborative fuzzy-neural approach for long-term load forecasting in Taiwan," *Computers & Industrial Engineering*, vol. 63, no. 3, pp. 663-670, 2012.

[14] K. Padmakumari, K.P. Mohandas, and S. Thiruvengadam, "Long term distribution demand forecasting using neuro fuzzy computations," *Electr Power Energy Syst.*, vol. 21, no. 5, pp. 315-322, 1999.

[15] W.C. Hong, "Electric load forecasting by support vector model," *Applied Mathematical Modelling*, vol. 33, pp. 2444-2454, 2009.

[16] W. Jianjun, L. Li, and L. Ding, "Application of SVR with backtracking search algorithm for long-term load forecasting," *J. Intel. Fuzzy Syst.*, vol. 31, no. 4, pp. 2341-2347, 2016.

[17] P.F. Pai and W.C. Hong, "Support vector machines with simulated annealing algorithms in electricity load forecasting," *Energy Conv. Management*, vol. 46, pp. 2669-2688, 2005.

[18] K. Karabulut, A. Alkan and A.S. Yilmaz, "Long term energy consumption forecasting using genetic programming," *Math Comput Appl*, vol. 13, no. 2, pp. 71-80, 2008.

[19] H.Z. Li, S. Guo, C.J. Li and J.Q. Sun, "A hybrid annual power load forecasting model based on generalized regression neural network with fruit fly optimization algorithm," *Knowledge-Based Systems*, vol. 37, pp. 378-387, 2013

[20] H. Abdi, and S.D Beigvand, "Long-term load forecasting based on gravitational search algorithm," *J. Intel. Fuzzy Syst.*, vol. 30, no. 6, pp. 3633-3643, 2016.

[21] A. Unler, "Improvement of energy demand forecasts using swarm intelligence: the case of Turkey with projections to 2025," *Energy Policy*, vol. 36, no. 6, pp. 1937-1944, 2008.

[22] M.R. AlRashidi and K.M. El-Naggar, "Long term electric load forecasting based on particle swarm optimization," *Applied Energy*, vol. 87, no. 1, pp. 320-326, 2010

[23] M. Meng and D. Niu, "Annual electricity consumption analysis and forecasting of China based on few observations methods," *Energy Conv. Management*, vol. 52, no. 2, pp. 953-957, 2011.

[24] J.W. Sun, "Energy demand in the fifteen European Union countries by 2010- a forecasting model based on the decomposition approach," *Energy*, vol. 26, pp. 549-560, 2001.

[25] J. Moral-Carcedo and J. Pérez-García, "Integrating long-term economic scenarios into peak load forecasting: An application to Spain," *Energy*, vol. 140, pp. 682-695, 2017.

[26] D. Angelopoulos, J. Psarras, and Y. Siskos, "Long-term electricity demand forecasting via ordinal regression analysis: The case of Greece,". *in Proc. IEEE PowerTech,* Manchester, 2017.

[27] Z. Yumurtaci and E. Asmaz, "Electric energy demand of Turkey for the year 2050," *Energy Source*, vol. 26, pp. 1157-1164, 2004.

[28] Z. Dilaver and L.C. Hunt, "Turkish aggregate electricity demand: an outlook to 2020," *Energy*, vol. 36, pp. 6686-6696, 2011.

[29] C. Hamzacebi and H.A. Es, "Forecasting the annual electricity consumption of Turkey using an optimized grey model," *Energy*, vol. 70, pp. 165-171, 2014.

[30] S.H.A. Kaboli, A. Fallahpour, J. Selvaraj, and N.A. Rahim, "Long-term electrical energy consumption formulating and forecasting via optimized gene expression programming," *Energy*, vol. 126, pp. 144-164, 2017

[31] H. Zareipour, K. Bhattacharya and C. Canizares, "Electricity market price volatility: The case of Ontario," *Energy Policy*, vol. 35, pp. 4739–4748, 2007.

[32] M. Benini, M. Marracci, P. Pelacchi and A. Venturini, "Day-ahead market price volatility analysis in deregulated electricity markets," Proc. 2002 IEEE Power Engineering Society Summer Meeting, Chicago, IL, USA.

[33] B.R. Auer, "How does Germany's green energy policy affect electricity market volatility? An application of conditional autoregressive range models," *Energy Policy*, vol. 98, pp. 621-628, 2016.

[34] J. Boland, J.A. Filar, G. Mohammadian and A. Nazari, "Australian electricity market and price volatility," *Annals of Operations Research*, vol. 241, no. 1-2, pp. 357-372, 2016

[35] Y. Li and P.C. Flynn, "Deregulated power prices: Comparison of volatility," *Energy Policy*, vol. 32, pp. 1591-1601, 2004.

[36] I. Simonsen, "Volatility of power markets," *Phys. Stat. Mech. Appl.*, vol. 335, pp. 10-20, 2005.

[37] S.R. Khuntia, J.L. Rueda, and M.A.M.M. van der Meijden, "Volatility in electrical load forecasting for long-term horizon—An ARIMA-GARCH approach," *in Proc. IEEE PMAPS*, Beijing, 2016.

[38] R.F. Engle, "New frontiers for ARCH models," *J. Applied Econometrics*, vol. 17, no. 5, pp. 425-446, 2002.

[39] R.F. Engle, and J.R. Russell, "Autoregressive conditional duration: A new model for irregularly spaced transaction data," *Econometrica*, vol. 66, pp. 1127–1162, 1998.

[40] M. Lanne, "A mixture multiplicative error model for realized volatility," *J. Financial Econometrics*, vol. 4, no. 4, pp. 594-616, 2006.

[41] H. Han, M.D. Park, and S. Zhang, "A multiplicative error model with heterogeneous components for forecasting realized volatility," *J. Forecasting*, vol. 34, no. 3, pp. 209-219, 2015.

[42] M. Caporin, E. Rossi, and P.S. de Magistris, « Chasing volatility: A persistent multiplicative error model with jumps," *J. Econometrics*, vol. 198, no. 1, pp. 122-145, 2017.

[43] G.E. Box, G.M. Jenkins, and G.C. Reinsel. *Time series analysis: forecasting and control.* John Wiley & Sons, Hoboken, 2011.

[44] Weblink: http://www.pjm.com/markets-and-operations/ops-analysis/

[45] National Bureau of Economic Research (NBER), *Business Cycle Expansions and Contractions*, http://www.nber.org/cycles/cyclesmain.html.

[46] J. Elder, and P.E. Kennedy, "Testing for unit roots: What should students be taught?," *J. Economic Education*, vol. 32, pp. 137–146, 2001.

[47] A.C. Johnson Jr, M.B. Johnson, and R.C. Buse, *Econometrics: Basic and applied* (p. 90). New York, 1987.

[48] G.M. Ljung, and G.E. Box, "On a measure of lack of fit in time series models," *Biometrika*, pp. 297-303, 1978.

[49] K. Ahoniemi, "Modeling and forecasting implied volatility - an econometric analysis of the VIX index.," *HECER Discussion Paper, 129*, 2006.

[50] T. Bollerslev, "Generalized autoregressive conditional heteroskedasticity," *J. Econometrics*, vol. 31, pp. 307-327, 1986.

[51] C. Brooks, and M.C. Oozeer, "Modelling the implied volatility of options on long gilt futures," *J. Business Finance & Accounting,* vol. 29, no. 1-2, pp. 111-137, 2002.

[52] P.R. Hansen, and A. Timmermann. *Choice of sample split in out-of-sample forecast evaluation*. Working paper, Stanford University, 2012.

[53] C.M. Jarque, and A.K. Bera. "A test for normality of observations and regression residuals," *International Statistical Review*, vol. 55, no. 2, pp. 163–172, 1987.

[54]    N.N. Taleb, "Black swans and the domains of statistics," *The American Statistician*, vol. 61, no. 3, pp. 198-200, 2007.